# Efficient Construction of Neighborhood Graphs by the Multiple Sorting Method

**Takeaki Uno**                                                          UNO@NII.JP
*National Institute of Informatics, 2-1-2, Hitotsubashi, Chiyoda-ku, Tokyo, Japan*

**Masashi Sugiyama**                                              SUGI@CS.TITECH.AC.JP
*Department of Computer Science, Tokyo Institute of Technology*
*2-12-1 O-okayama, Meguro-ku, Tokyo 152-8552, Japan*

**Koji Tsuda**                                                   KOJI.TSUDA@AIST.GO.JP
*Computational Biology Research Center, National Institute of Advanced Industrial Science and Technology (AIST) 2-42 Aomi Koto-ku, 135-0064 Tokyo, Japan*

**Editor:** Leslie Pack Kaelbling

## Abstract

Neighborhood graphs are gaining popularity as a concise data representation in machine learning. However, naive graph construction by pairwise distance calculation takes $O(n^2)$ runtime for $n$ data points and this is prohibitively slow for millions of data points. For strings of equal length, the multiple sorting method (Uno, 2008) can construct an $\epsilon$-neighbor graph in $O(n+m)$ time, where $m$ is the number of $\epsilon$-neighbor pairs in the data. To introduce this remarkably efficient algorithm to continuous domains such as images, signals and texts, we employ a random projection method to convert vectors to strings. Theoretical results are presented to elucidate the trade-off between approximation quality and computation time. Empirical results show the efficiency of our method in comparison to fast nearest neighbor alternatives.

## 1. Introduction

Neighborhood graphs are widely used in various machine learning and data mining tasks such as manifold modeling (Tennenbaum et al., 2000), semi-supervised learning (Zhou et al., 2004), spectral clustering (Hein et al., 2007), and retrieval of protein sequences (Weston et al., 2004). There are two types of neighborhood graphs, $k$-nearest neighbor graphs (a node is connected to its $k$ nearest neighbors) and $\epsilon$-neighbor graphs (two nodes whose distance is within $\epsilon$ are connected). Naive construction of neighborhood graphs take $O(n^2)$ distance calculations, where $n$ denotes the number of data points. This is prohibitively slow in recent large-scale applications. This paper deals with fast methods to create $\epsilon$-neighbor graphs. To be precise, let us define our problem as follows.

**Definition 1.1 ($\epsilon$-Neighbor Graph Construction)** *Given a set of vectors $\boldsymbol{x}_1, \ldots, \boldsymbol{x}_n \in \Re^d$, enumerate all pairs $(i,j), i < j$ such that $\Delta(\boldsymbol{x}_i, \boldsymbol{x}_j) \leq \epsilon$, where $\Delta$ is a metric such as Euclidean distance and cosine distance.*

The problem of $\epsilon$-neighbor graph construction is related to, but distinctly different from the $\epsilon$-neighbor search.

**Definition 1.2 ($\epsilon$-Neighbor Search)** *Given a query point $\boldsymbol{v}$ and a set of vectors, enumerate all neighbors $\boldsymbol{x}_i$ satisfying $\Delta(\boldsymbol{v}, \boldsymbol{x}_i) \leq \epsilon$.*

The problem of $\epsilon$-neighbor graph construction is simpler than $\epsilon$-neighbor search, because any $\epsilon$-neighbor search algorithm can solve $\epsilon$-neighbor graph construction by applying the algorithm to every point, but not vice versa. Conventionally, however, the neighborhood graphs have been constructed by nearest neighbor search methods (Datar et al., 2004; Beygelzimer et al., 2006; Omohundro, 1991). There are two kinds of methods, exact and approximate. Exact methods including cover tree (Beygelzimer et al., 2006), ball tree (Omohundro, 1991) and kd-tree (Lee and Wong, 1977) can find all neighbors without fail. Approximate methods such as E2LSH (Datar et al., 2004) use random projections called locality sensitive hashing (LSH) to classify the points into small cells. Neighbors of a query point are sought only in the cell that includes the query point. Approximate methods can miss a small fraction of neighbors, but they are theoretically faster than the exact alternatives. In practical problems, however, such approximate methods do not compare favorably with exact methods with respect to running time, see e.g., (Beygelzimer et al., 2006).

Our approach aims to solve the neighbor graph construction problem *directly* by the following two steps. First, data points are mapped to strings of discrete symbols using LSH. Through this mapping, the closeness between two vectors is approximately preserved as the Hamming distance between strings. Subsequently, all pairs whose Hamming distance is within $d$ are enumerated by the multiple sorting method (MSM) (Uno, 2008), which is an exact enumeration method based on masked radix sort. Since our algorithm includes stochasticity in the random projection part, missing edges can occur in the neighborhood graph with a certain probability. Nevertheless our algorithm can explicitly control the expected fraction of missing edges (*missing edge ratio*) under a small value, e.g., $1.0 \times 10^{-6}$.

Since our method uses LSH, one might worry that the practical slowness of LSH-based neighbor search carries on in our method. However, the crucial difference is that the existing LSH methods try to accomplish the search by random projection alone. Namely, one has to design the projection such that neighbors are mapped to the same string exactly. In general, it is hard to create single mapping to achieve this. In common practice, many replicates of strings are used to boost the true discovery rate (Datar et al., 2004). Such a scheme results in very redundant strings: Typically, more than 100 replicates of relatively short strings (length 10-50) are used to reach a reasonable level of accuracy (Andoni and Indyk, 2005). See (Weiss et al., 2009) for related discussions. On the other hand, in our method, we design the projection such that neighbors are mapped to *similar* strings, not identical (more precisely, the Hamming distance is smaller than $d$). So, the number of necessary replicates is typically much smaller than the existing LSH-based neighbor search.

In empirical evaluation based on 1.6 million images (Torralba et al., 2008), our method was significantly faster than the cover tree while keeping the missing edge ratio under $1.0 \times 10^{-6}$. It will be shown that good efficiency cannot be attained, if we try to solve the problem by LSH alone.

The rest of this paper is organized as follows. Section 2 reviews the multiple sorting method for strings. In Section 3, locality sensitive hashing is introduced and the drawbacks of existing LSH-based neighbor search are discussed. In Section 4, we present our approach and evaluate the missing edge ratio. Section 5 presents empirical results. Section 6 presents further discussion and closing remarks.

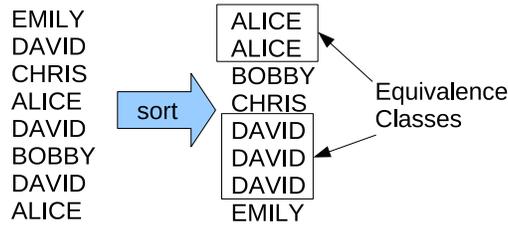

Figure 1: Sorting and equivalence classes.

## 2. Multiple Sorting Method

In this section, we review the multiple sorting method (MSM) (Uno, 2008).

### 2.1 Basic Idea

Consider the problem of constructing the $d$-neighbor graph of string data of equal length $\ell$. As the metric, we use the Hamming distance, i.e., the number of disagreements. The problem is formulated as follows: Given a string pool $S = \{s_1, \ldots, s_n\}$, find all pairs $(i, j)$, $i < j$ whose Hamming distance is at most $d$, $HamDist(s_i, s_j) \leq d$. The multiple sorting method (MSM) (Uno, 2008) can solve the problem in $O(n + m)$ time, where $m$ is the number of $d$-neighbor pairs,

$$m = |\{(i, j) \mid HamDist(s_i, s_j) \leq d\}|. \tag{1}$$

To explain the idea of MSM intuitively, let us start from the special case $d = 0$, that is, enumerating exactly same string pairs. In that case, the problem is solved by sorting the strings and scanning the sorted list to divide it into equivalence classes (Figure 1). Then, for each equivalence class, edges are built between all pairs. Using radix sort, sorting takes only $O(n)$ time. The edge building takes $O(m)$ time. So the overall complexity is $O(n+m)$.

Even if $d > 0$, we can enumerate neighbor pairs by applying radix sort multiple times. Let $C$ denote a set of $d$ distinct integers taken from $\{1, \ldots, \ell\}$. Denote by $s_i^C$ the $i$-th string whose characters at positions $C$ are removed. We call them $C$-masked strings. Obviously, the following two statements are equivalent.

- There exists $C$ such that $s_i^C = s_j^C$, $|C| = d$.

- $HamDist(s_i, s_j) \leq d$.

Therefore, the neighbor pairs can be enumerated by trying every possible $C$ of size $d$ and sorting the masked strings. It takes $\binom{\ell}{d}$ times sorting, hence the time complexity is polynomial to $\ell$ and exponential to $d$. Nevertheless, in terms of $n$ and $m$, the time complexity stays linear, yielding overall complexity $O(n + m)$. Figure 2a and 2b illustrate multiple sorting with masks.

Figure 2: Multiple sorting method.

## 2.2 Blockwise Masking

Although the above method is optimal in terms of complexity, the practical computation time is not always optimal because a large number of sorting operations are necessary. To reduce the number of sorting operations, *blockwise masking* comes in useful.

Let us divide the strings into $k$ blocks of approximately equal length as in Figure 2c. Define $B$ as a set of $d$ distinct integers taken from $\{1, \ldots, k\}$. Denote by $s_i^B$ the $i$-th string whose blocks listed in $B$ are removed. If $HamDist(s_i, s_j) \leq d$, then there exists $B$ such that $s_i^B = s_j^B$. However, the inverse does not hold. When pairs are enumerated by trying every possible $B$ and sorting the masked strings as before, the solution set contains all neighbor pairs as well as a certain number of non-neighbor pairs. To filter out non-neighbor pairs, we need to calculate the actual Hamming distances. Since distance calculation is done only for pairs falling into an equivalence class, the number of distance calculations is much smaller than exhaustive comparison. Indeed, the number of sorting operations is significantly reduced: from $\begin{pmatrix} \ell \\ d \end{pmatrix}$ to $\begin{pmatrix} k \\ d \end{pmatrix}$. In the example shown in Figure 2c, the number of sorting operations is reduced from 120 to 6. Due to space limitation, we cannot write full details of the computational tricks. For further description, see Appendix and the original paper (Uno, 2008).

### 2.3 Output Sensitive Complexity

The computational complexity of MSM depends on the number of solutions $m$. Such complexity is called *output sensitive* (Johnson et al., 1988). Sometimes, the complexity depending on the input only is not a good description of reality. For example, the input-only complexity of MSM is $O(n^2)$, which is evaluated in the worst case that all strings are identical. One drawback of output sensitive complexity is that it cannot be used for foreseeing the computational time before running the algorithm. However, it is very useful for efficiency comparison among algorithms.

## 3. Locality Sensitive Hashing

In this section, we review existing locality sensitive hashing (LSH) methods for nearest neighbor discovery.

### 3.1 Basic Idea

Locality sensitive hashing is a random mapping from a vector to a string of integers, $\Re^D \to \mathcal{I}^\ell$. The data points are mapped to the strings and stored in a hash table. When a query point is given, it is converted to a string, which is then used as a key to retrieve the data points with the same key. Finally, the retrieved points whose actual distances are within $\epsilon$ are reported. It is known that a single mapping cannot control the number of false negatives (i.e., neighbors with non-identical keys) (Datar et al., 2004). Therefore, it is common to create multiple hash tables and use them simultaneously. Historically, the first LSH algorithm has been proposed for the Hamming space (Gionis et al., 1999). Later, it has been extended to the Euclidean distance (Datar et al., 2004). A similar mapping for the cosine distance was originally used in approximating the max-cut problem (Goemans and Williamson, 1995). In the following, we focus on cosine LSH that is actually used in our experiments. Notice, however, that any kind of LSH can be combined with MSM in principle.

### 3.2 Cosine LSH

Denote $n$ data points in $\Re^D$ by $\boldsymbol{x}_1, \ldots, \boldsymbol{x}_n$. Let us take the cosine distance as our metric.

$$\Delta(\boldsymbol{x}_i, \boldsymbol{x}_j) = 1 - \frac{\boldsymbol{x}_i^\top \boldsymbol{x}_j}{\|\boldsymbol{x}_i\|\|\boldsymbol{x}_j\|}. \tag{2}$$

We would like to enumerate all pairs whose cosine distance is at most $\epsilon$.

Let $R \in \Re^{D \times \ell}$ be a random matrix consisting of i.i.d. samples from the standard normal distribution $N(0, 1)$. Let $\boldsymbol{s}_1, \ldots, \boldsymbol{s}_n$ be bit strings of length $\ell$ consisting of '0' and '1'. The projection is defined as

$$s_{ik} := \text{sign}(\boldsymbol{r}_k^\top \boldsymbol{x}_i), \tag{3}$$

where $s_{ik}$ is the $k$-th character of the $i$-th string, $\boldsymbol{r}_k$ is the $k$-th column of $R$ and $\text{sign}(t)$ produces the character '1' if $t > 0$ and '0' otherwise. It is known (Broder et al., 2000) that the following relationship holds:

$$\Pr\left(s_{ik} \neq s_{jk}\right) = \frac{\theta_{ij}}{\pi}, \quad \forall k, \tag{4}$$

where $\theta_{ij}$ is the angle between $\boldsymbol{x}_i$ and $\boldsymbol{x}_j$:

$$\theta_{ij} = \arccos\left(\frac{\boldsymbol{x}_i^\top \boldsymbol{x}_j}{\|\boldsymbol{x}_i\|\|\boldsymbol{x}_j\|}\right).$$

This relationship guarantees that the expected value of $HamDist(\boldsymbol{s}_i, \boldsymbol{s}_j)$ is a monotonically increasing function of the cosine distance. Furthermore, in the limit $\ell \to \infty$, the Hamming distance between two bit strings converge to the angle of the original vectors.

$$\lim_{\ell \to \infty} HamDist(\boldsymbol{s}_i, \boldsymbol{s}_j) = \frac{\theta_{ij}}{\pi}.$$

When $\ell$ is finite, $HamDist(\boldsymbol{s}_i, \boldsymbol{s}_j)$ is subject to the binomial distribution $\mathrm{Binom}(L, \frac{\theta_{ij}}{\pi})$.

## 4. Multiple Sorting Method for Continuous Data

This section presents our method for neighborhood graph construction.

### 4.1 Basic Idea

The basic idea is to map the data points to strings by LSH, and enumerate pairs of similar strings by MSM. However, considering the fact that the computational complexity of MSM is a polynomial function of the string length, it is not a good strategy to create long strings and process them by MSM at once. Thus, we employ the following replication strategy: for relatively small $\ell$, we create strings of length $\ell$ from the data points by LSH. It is repeated $Q$ times, resulting in $Q$ independent string pools. Denote by $\boldsymbol{s}_i^q$ the $q$-th string corresponding to $\boldsymbol{x}_i$. Then, MSM is applied to each string pool. The $q$-th output set of MSM is described as

$$E_q = \{(i,j) \mid HamDist(\boldsymbol{s}_i^q, \boldsymbol{s}_j^q) \le d, i < j\}. \tag{5}$$

Finally, the output sets are merged into one,

$$E = E_1 \cup \cdots \cup E_Q.$$

Then, we compute actual distances $\Delta(\boldsymbol{x}_i, \boldsymbol{x}_j)$ for all pairs in $E$, and report the pairs that satisfy $\Delta(\boldsymbol{x}_i, \boldsymbol{x}_j) \le \epsilon$.

### 4.2 Missing Edge Ratio

Given the true edge set $E^*$ and our intermediate solution $E$, there are two kinds of error.

- Type-I error (false positive): A non-neighbor pair has a Hamming distance within $d$ in at least one replicate.

$$F_1 = \{(i,j) \mid (i,j) \in E, (i,j) \notin E^*\}.$$

- Type-II error (false negative): A neighbor pair has a Hamming distance larger than $d$ in all replicates.

$$F_2 = \{(i,j) \mid (i,j) \notin E, (i,j) \in E^*\}.$$

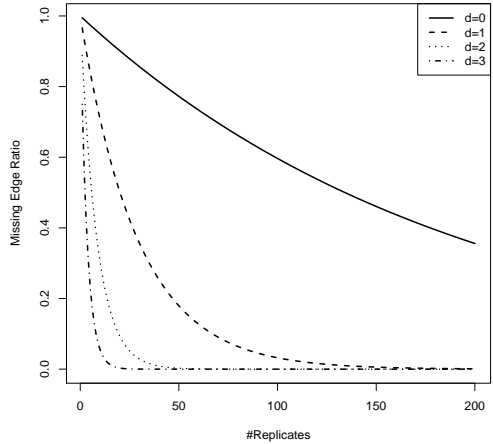

Figure 3: The bound of missing edge ratio against the number of replicates.

The type-II errors are more critical in our method because the type-I errors are eventually filtered out by calculating the actual distances in the post-processing step. However, introducing too many type-I errors harms computational efficiency because $|E|$ distance calculations are necessary.

The fraction of missing edges is defined as $|F_2|/|E^*|$, whose expectation (*missing edge ratio*) is bounded as follows.

$$E\left[\frac{|F_2|}{|E^*|}\right] \leq \left(1 - \sum_{k=0}^{\lfloor d \rfloor} \left(\begin{array}{c} \ell \\ k \end{array}\right) p^k (1-p)^{\ell-k}\right)^Q,$$

where $p$ is an upper bound of the non-collision probability (4) for neighbors. For the cosine LSH, $p$ is set as follows,

$$p = \frac{\arccos(1 - \epsilon)}{\pi}.$$

There is no way to derive non-trivial type-I error bounds without the knowledge of data distribution. However, we found in empirical evaluations that the type-I error is typically small enough to keep our method competitive against other methods.

Figure 3 depicts the missing edge ratio as a function of $Q$ for different values of $d$. We used the cosine LSH where the radius is set such that $p = 0.1$ and $L = 50$. It is observed that many replicates are required to achieve small missing edge ratio when $d = 0$. This plot illustrates the difficulty of performing nearest neighbor search by hashing alone. As $d$ increases, the number of required replicates reduces remarkably.

### 4.3 Setting Parameters

Our method has three parameters $(d, \ell, Q)$. Users specify the radius $\epsilon$ and the upper bound of the missing edge ratio $\gamma$. We need to specify the parameters to reduce the number of type-I and type-II errors as much as possible. Basically, we first choose $\ell$, then $d$, and finally choose the minimum $Q$ such that the the missing edge ratio is smaller than $\gamma$. Towards the efficient computation, we observe the following.

- Strings generated by random projection should be short. approximately, the computation time of MSM linearly increases against the string length.

- The length $\ell$ should also be short, but this increases the number of type-I errors.

- The number of replicates $Q$ should be small.

- The number of mismatches $d$ should be small. The computation time of MSM increases exponentially against the increase of $d$.

Of course they conflict, thus we need to choose good parameters in some way. Actually, $Q$ grows exponentially against the decrease of $d$, thus the computation time will be long if $d$ is too small or too large, and the best choice of $d$ should be the middle. In our pre-experiments with fixed $\ell$ and changed $d$ and $Q$, when the number of replicates is large, say larger than 25, the increase $d$ by one usually decreases the computation time. Thus, in our parameter setting, we first choose $\ell$, then choose the smallest $d$ such that $Q$ will be less than 25.

In the computation of MSM, MSM chooses $k - d$ blocks and sort the strings according to the blocks. Thus, we should have sufficiently many letters in one block so that the cost for pairwise comparison in the groups of strings having the same blocks. For the purpose, $k - d$ blocks should have at least $\log_2 n$ letters since the expected number of vectors having the same blocks will no greater than 1. Thus, we set $\ell$ to $2 \log_2 n$. This ensures that the efficiency of MSM for small $d$. For large $d$ such as 10, MSM actually does not perform well, but in such a case even if we take large $\ell$, the performance is bad. This occurs when the threshold value is large, and there may be up to $\Theta(n^2)$ close pairs. For such problems, straightforward pairwise comparison is the best choice.

If the output pairs are few, type-I errors are also few, usually. In such cases, especially if $n$ is quite large, The bottle neck of the computation is always the generation of the strings to be compared, because it involves huge number of inner products. In such cases, we should reduce $\ell$ and increase $d$ so that the computation time of MSM and random projection will be the same. This needs the estimation of the number of output pairs. In our method, we randomly sample the vector pairs, and roughly estimate the output pairs. Let $S$ be the number of estimated output. In our method, we limit the number of replicates to $\max\{30S/n, 5\}$. The number 30 is chosen by pre-experiments; usually, type-I errors are no greater than 30 times more closed pairs. We also slightly shorten $\ell$ according to $S$; when $S$ is closed to 0, we decrease $\ell$ to half.

For the E²LSH, we have one more choice, the width parameter $w$. The probability of that two vectors have the same hash value is small if $w$ is small, and large if $w$ is large. Thus, basically smaller $w$ makes $d$ larger, and large $w$ involves many type-I errors. Thereby, we also use estimated output size $S$, and choose large $w$ if $S$ is small. In our pre-experiments with quite small threshold distance, $w = 12$ was the best. With large threshold, $w = 3.5$ was the best. Type-I errors seems to increase mildly exponentially as the increase of $w$, thus according to those experiments, we considered that output pairs are many if $S > 10n$, and we set $w$ to $3.5 - \frac{\log(S/10n)}{4}$. We set $w$ to 3.5 if $w$ is smaller than 3.5, and set $w$ to 12 if $w$ is larger than 12.

## 5. Experiments

We here present the results of computational experiments to evaluate the practical efficiency. All experiments are executed in a Linux PC with Intel Core2Duo E8400 of 3.0GHz with 4GB memory.

### 5.1 Methods in Comparison

Our method is compared with the cover tree (Beygelzimer et al., 2006), a state-of-the-art exact nearest neighbor method. In the original paper, they perform k-nearest neighbor search only, not $\epsilon$-neighbor search. However, the cover tree can be used for both purposes in principle. We downloaded the template code, `http://hunch.net/~jl/projects/cover_tree/cover_tree_2.tar.gz`, and used the function `epsilon_nearest_neighbor()` developed for $\epsilon$-neighbor search. The cover tree is based on the Euclidean distance, while our method is based on the cosine distance (2). For fair comparison, vectors are centralized and normalized to have norm one, under which the squared Euclidean distance is equivalent to the doubled cosine distance. The radius for the Euclidean distance is set as $\sqrt{2\epsilon}$ where $\epsilon$ is the radius for the cosine distance.

In addition, we evaluated a valiant of our method where the distance threshold $d$ is set to zero. Since this method essentially uses the locality sensitive hashing only, it is referred to as "LSH" below. The parameter setting procedure presented in Section 4.3 is not appropriate for LSH, because, at a small threshold of $\gamma$, the number of required replicates gets very large. So, we fixed the number of replicates $Q = 300$ a priori, and adjusted the string length $\ell$ to achieve the required level of type-II error.

### 5.2 Efficiency Results

The dataset we used is the set of small images collected by (Torralba et al., 2008). The dataset has 8 million images in total, but we used a smaller version containing 1.6 million images, which was immediately downloadable from `http://people.csail.mit.edu/torralba/tinyimages/`. Images are converted to gray scale, yielding 1024 dimensional vectors. To observe the growth rate, we generated data of different sizes by taking the first $n$ records. In neighborhood graph construction, we tried three different values of cosine distance radius $\epsilon = 0.0123, 0.0489$ and $0.109$ which translate to $0.05\pi$, $0.10\pi$ and $0.15\pi$ in terms of angle, respectively. It seems to be meaningless to try larger thresholds, because the number of edges reached 0.2% of all pairs when $\epsilon = 0.109$ and $n = 320000$. It means that each node has 640 neighbors on average. To make the number of missing edges negligibly small, the missing edge ratio bound was set to $\gamma = 1.0 \times 10^{-6}$ for MSM and LSH.

Figure 4 shows the efficiency results. Our MSM method was consistently more efficient than the cover tree. Already in small data sizes ($\approx 40000$), the difference is clear. The time of cover tree contains both tree construction and the searches over the tree. One important thing to notice, however, is that MSM cannot process new query points, while the cover tree can. The cover tree spends much time to prepare a shallow search tree to enable fast search for unknown query points. Our algorithm does not need to construct such a tree, because our task is limited to the data points at hand.

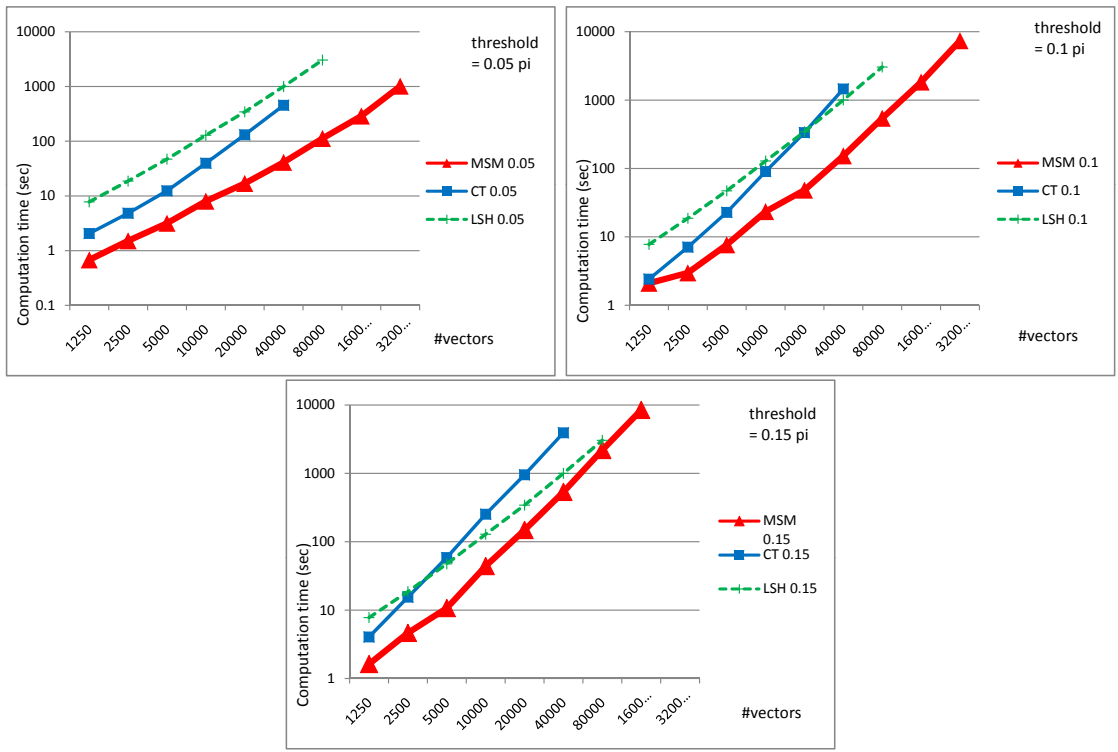

Figure 4: Comparison of computation time.

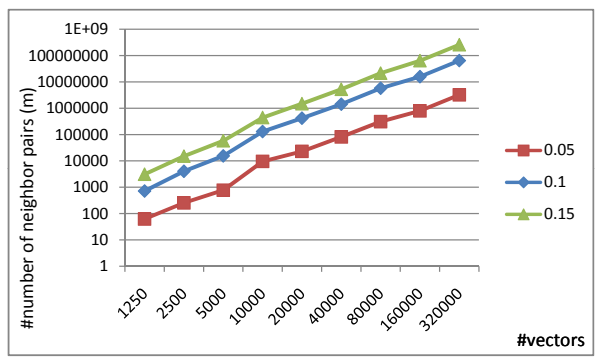

Figure 5: Comparison of the number of neighbor pairs $m$.

Figure 5 shows the comparison of the number of neighbor pairs $m$. This shows that irrespective of the threshold, the number of neighbor pairs $m$ gets approximately three times larger when the number of vectors $n$ is doubled. This implies that $m$ is a super-linear function of $n$; this phenomenon is essentially different from $k$-nearest neighbor graph construction in which $m$ is linear in $n$.

Table 1: Computational Time of MSM for 1.6 million images.

| Angle Threshold | #neighbor pairs | time(sec) |
|---|---|---|
| $0.01\pi$ | 3382 | 1409 |
| $0.02\pi$ | 36881 | 3277 |
| $0.03\pi$ | 334462 | 4391 |
| $0.04\pi$ | 1533622 | 7525 |

The computational results on 1.6 million images are separately summarized in Table 1. At the angle threshold $0.04\pi$, MSM finishes in around 2.09 hours creating around 1.5 million edges. It corresponds to around 1.9 neighbor per node.

### 5.3 Empirical Missing Edge Ratio

Using the dataset with 160000 points, we plotted the empirical ratio of missing edges in Figure 6. This is not a theoretical bound, but actual counts. As expected from the bound, the missing ratio is reduced exponentially as the number of replicates increases. In the experiments of Figure 4, we set the threshold to $1.0 \times 10^{-6}$, but it is quite easy to impose an even stricter threshold without large loss of efficiency.

### 5.4 Output Sensitivity

Output sensitive methods can finish the task quickly if the number of solutions is small. In our problem, MSM finishes quickly if $\epsilon$ is small. Not all algorithms have this property. For example, the naive pairwise calculation spends the same amount of time regardless of $\epsilon$. This property is guaranteed theoretically for MSM, but we would like to see that MSM has output sensitivity empirically as well. Figure 7 shows the computation time per 10,000 output pairs. The curve is flat or slightly decreasing, showing that MSM is output sensitive in fact.

## 6. Discussion and Concluding Remarks

In this paper, we started by defining the problem of neighborhood graph construction and characterizing the difference from neighbor search. We have shown that our method, an extension of MSM for continuous domains, can be more efficient than the cover tree with a negligible ratio of missing edges. Here we combined MSM with random projections, but it is straightforward to combine MSM with other non-random hashing methods such as spectral hashing (Weiss et al., 2009) and semantic hashing (Salakhutdinov and Hinton, 2007). Such methods put emphasis on preserving important statistical structure rather than mapping similar vectors to similar strings.

The idea of using sorting operations for finding neighbors initiated by (Uno, 2008) is quite stimulating. The topic of this paper was the combination of MSM and random projections, but this is not the end of story for us. In future work, we would like to explore the following questions. 1) Is it really necessary to map the vectors to discrete symbols? Apart from locality sensitive hashing, there are a variety of random projection algorithms that maps a vector to a real-value (Li et al., 2006). Can they be used for better efficiency?

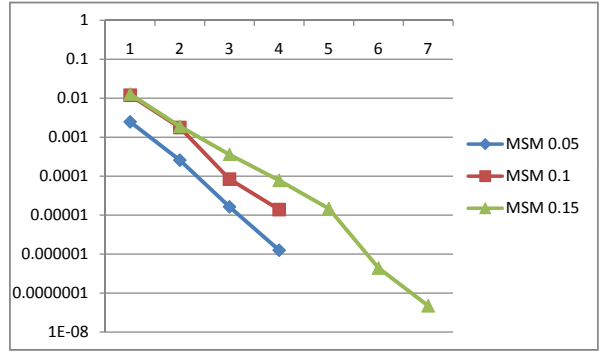

Figure 6: Empirical ratio of missing edges. There were no missing edges when the number of replicates $Q \geq 5$ and the angle threshold is $0.1\pi$ and $0.05\pi$.

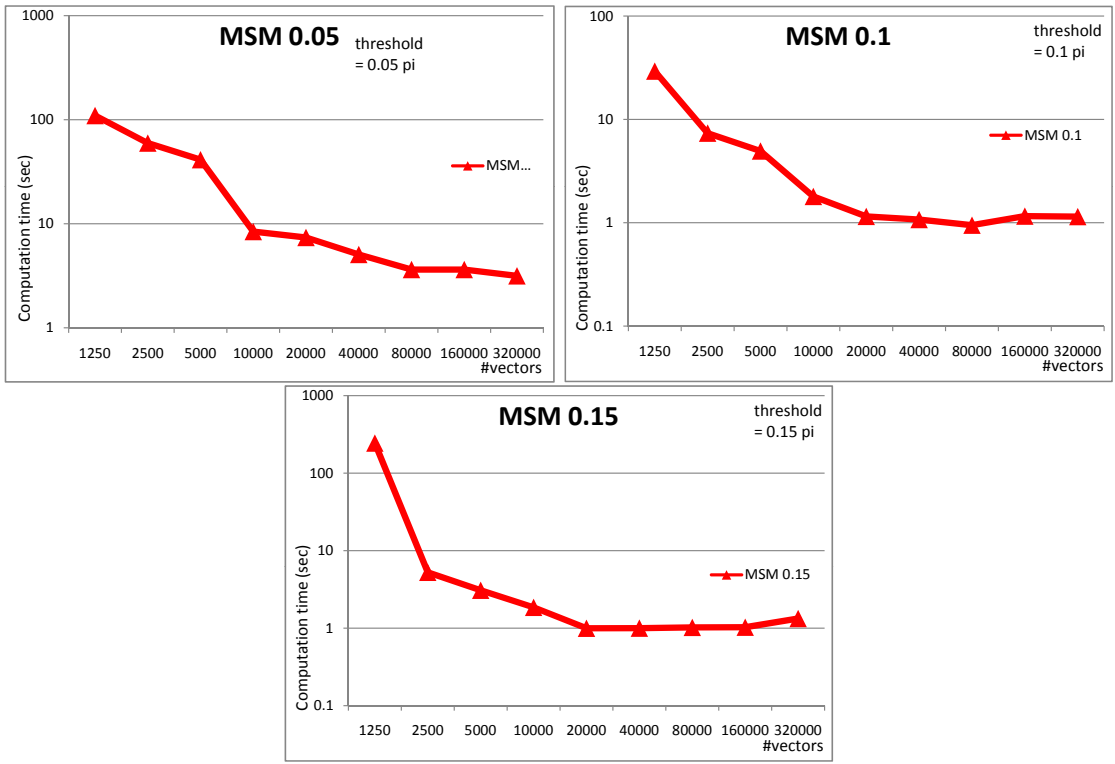

Figure 7: Computation time per 10,000 output pairs.

Sorting continuous values takes more time than radix sort, but it is possible in $O(n \log n)$ time. 2) Hierarchical organization of the multiple sorting method. Our current algorithm has a two-level structure that $Q$ replicates are created, each of which is divided into $k$ blocks. We are not yet sure if it is the optimal architecture. Further efficiency could possibly be achieved by organizing the algorithm in a hierarchy of more than two levels. 3) Implementation of MSM in a many-core processor. Though we performed all sorting operations serially, MSM is inherently amenable to parallelization. In the coming era of many-core processors and massive parallelism (Manferdelli et al., 2008), MSM is highly promising as a standard method for neighborhood graph construction. Fortunately, radix sort is available in most GPU libraries. It would be interesting to implement MSM on a GPU and see how fast it can be.

## Appendix: Acceleration Tricks for MSM

In actual implementation of MSM, some computational tricks are effective in reducing the computational cost. In this section, we briefly describe some of them.

The first trick is for checking duplications in $E_1, \ldots, E_Q$. In taking the union of them $E = E_1 \cup \cdots \cup E_Q$, we do the following. Suppose that $Q$ strings $\boldsymbol{s}_j^q, \ldots, \boldsymbol{s}_j^Q$ are generated for the $j$-th vector by random projections. Even if we find a pair $\boldsymbol{s}_i^h$ and $\boldsymbol{s}_j^h$ having a Hamming distance at most $d$ for some $h$, we do not immediately output it to $E$. In that case, we compare $\boldsymbol{s}_i^g$ and $\boldsymbol{s}_j^g$ for all $1 \le g < h$, and, only if none of the compared pairs has the distance at most $d$, $(\boldsymbol{s}_i^h, \boldsymbol{s}_j^h)$ is added to $E$. In this way, we can ensure that no duplication happens.

The second trick is about the speed-up of radix sort. The time complexity of the radix sort is $O((n+\Sigma)\ell)$, where $\Sigma$ is the size of alphabet used in the input strings. In our problem, the alphabet is $\{0, 1\}$, thus $|\Sigma| = 2$. As it is very small compared to $n$, it is possible to save the cost by unifying several letters into one letter. For example, we unify every 20 letters into one letter so that the size of alphabet becomes $2^{20}$. This reduces the number of iterations in a radix sort from $\ell$ to $\lceil \ell/20 \rceil$. Radix sort inserts each string to one of $\Sigma$ buckets, then scan all the buckets in the increasing order of letters. In the case with large $|\Sigma|$, we can reduce the computation time by scanning only non-empty buckets. It means that we do not need to sort the strings fully, but only to order them such that the same strings appear consecutively. This reduces the time complexity from $O((n+\Sigma)\ell)$ to $O(n\ell)$, except for initialization. Initialization is not problematic, because it is performed only once, while radix sort is done many times.